\documentclass[fleqn,twoside]{article}
\usepackage{espcrc2}

\usepackage[figuresright]{rotating}

\newcommand{\AmS}{{\protect\the\textfont2
  A\kern-.1667em\lower.5ex\hbox{M}\kern-.125emS}}

\title{Braneworld Cosmology, the CMB and the Radion}

\author{A.C. Davis,\address{DAMTP, Centre for Mathematical Sciences,\\
        Cambridge University, \\
        Wilberforce Road, Cambridge, CB3 0WA, UK}
        Ph. Brax\address{Service de Physique Theorique, CEA-Saclay, \\
        F-91191, Gif/Yvette cedex, France}
        and
        C. van de Bruck\address{Department of Applied Mathematics \\
        Sheffield University, Sheffield, S3 7RH, UK}}

\begin{document}

\begin{abstract}
Recent developments in the theory of extra dimensions have opened
up avenues to confront such theories with cosmological tests. We
discuss a brane-world model with a bulk scalar field, motivated by
supergravity. The low-energy effective action is derived and
physical constraints on the parameters of the model discussed. The
cosmological evolution of the brane-world moduli is investigated
and it is shown that one of the moduli is a quintessence field.
The CMB predictions are computed. Finally, the possibility that
the radion field in brane-worlds could be a chameleon field is
investigated.

\end{abstract}

\maketitle

\section{Introduction}
A recent class of higher dimensional models have attracted the attention both
of particle physicists as well as cosmologists: brane world theories (see \cite{branereviews} 
for recent reviews).
The motivation comes
from developments in string theory, such as the discovery of
brane sources as well as the eleven--dimensional origin of the five
string theories in ten--dimensions. In particular the setup of
heterotic M--theory and its compacitification down to five dimensions \cite{lukas} 
leads to a well motivated five-dimensional brane world scenario, which
can be used (and generalized) to study its consequences in particle physics
and cosmology. We will concentrate on such setups in these proceedings.

Brane--world models in five dimensions have an exact low energy description
below the brane tension involving a four dimensional effective action.
Above the brane tension, non-conventional features such as the appearance
of the famous $\rho^2$ term, i.e. the
squared matter density, in the Friedmann equation precludes any attempt
to use a four dimensional approach \cite{langlois}. Below the brane tension, it has been
shown that the solutions of the five dimensional equations with boundary
terms are identical to the solutions of the four dimensional equations
derived from an effective action. This effective action is of the
scalar--tensor type with a universal coupling to matter on each brane.

In Section 2 we will present the moduli-space approximation
to derive the low energy
effective action and describe the coupling to matter. Post--Newtonian
constraints from solar system experiments will be used in Section 3
to restrict the coupling of the bulk scalar field to the brane
and the value of the radion today. Nucleosynthesis constraints are also
discussed. In Section 4 we will discuss the cosmological evolution of
brane-world moduli and show how exponential quintessence arises by
detuning the brane tension. In Section five we present the implication
of the moduli evolution for the cosmic microwave background anisotropies.
Finally, in Section 6, we apply the
so called chameleon mechanism to the radion and discuss its consequences.
Our conclusions are summarized in Section 7.

\section{The low-energy effective action}
We consider brane--world models in 5d with a bulk scalar field $\psi$. 
(We are following here \cite{brax1}, see e.g. \cite{bulkothers} for similar setups.) 
The bulk action consists of two terms which describe
gravity and the bulk scalar field dynamics:
\begin{eqnarray}
S_{\rm bulk} &=& \frac{1}{2\kappa_5^2} \int d^5 x \sqrt{-g_5}
\left( {\cal R} \right. \nonumber \\ 
&-& \left. \frac{3}{4}\left( (\partial \psi)^2 + U \right) \right).
\end{eqnarray}
Further, our setup contains two branes. One of these branes has a
positive tension, the other brane has a negative tension.  They are
described by the action
\begin{eqnarray}
S_{\rm brane 1} &=& -\frac{3}{2\kappa_5^2}\int d^5x \sqrt{-g_5} U_B
\delta(z_1), \label{b1} \\
S_{\rm brane 2} &=& +\frac{3}{2\kappa_5^2}\int d^5x \sqrt{-g_5} U_B \delta(z_2) \label{b2}.
\end{eqnarray}
In these expressions, $z_1$ and $z_2$ are the (arbitrary) positions of
the two branes, $U_B$ is the superpotential; $U$, the bulk potential
energy of the scalar field, is given by
\begin{equation}
U = \left(\frac{\partial U_B}{\partial \psi}\right)^2 - U_B^2.
\end{equation}
We will also include the Gibbons--Hawking boundary term for each
brane, which have the form
\begin{equation}
S_{\rm GH} = \frac{1}{\kappa_5^2}\int d^4 x \sqrt{-g_4} K,
\end{equation}
where $K$ is the extrinsic curvature of the individual branes.
We impose a $Z_2$--symmetry at the position of each brane.

The solution of the system above can be derived from
BPS--like equations of the form
\begin{equation}
\frac{a'}{a}=-\frac{U_B}{4},\ \psi'=\frac{\partial U_B}{\partial \psi},
\end{equation}
where $'=d/dz$ for a metric of the form
\begin{equation}\label{background}
ds^2 = dz^2 + a^2(z)\eta_{\mu\nu}dx^\mu dx^\nu.
\end{equation}
We will particularly focus on the case where the superpotential is an
exponential function:
\begin{equation}\label{potential}
U_B=4k e^{\alpha \psi}.
\end{equation}
The values  $\alpha =1/\sqrt 3,-1/\sqrt {12}$ were obtained in a theory
with supergravity in singular spaces \cite{brax1}. The solutions read
\begin{equation}\label{scale}
a(z)=(1-4k\alpha^2z)^{\frac{1}{4\alpha^2}},
\end{equation}
while the scalar field solution is
\begin{equation}\label{psi}
\psi = -\frac{1}{\alpha}\ln\left(1-4k\alpha^2z\right).
\end{equation}
In the $\alpha\to 0$ we retrieve the AdS profile
\begin{equation}
a(z)=e^{-kz}.
\end{equation}
Notice that in this case the scalar field decouples altogether. Also,
notice that there is a singular point in the bulk at
$z_* = 1/4k\alpha^2$, for which the scale factor vanishes.

In the following we will discuss the moduli space approximation \cite{moduliapprox}.
At low energy below the brane tension this leads to an exact description of the brane system.  Two of the moduli of the
system are the brane positions, i.e. in the solution above the
brane positions are arbitrary.  In the moduli space approximation,
these moduli are assumed to be space--time dependent.  We denote the
position of brane 1 with $z_1 = \phi(x^\nu)$ and the position of
brane 2 with $z_2 = \sigma(x^\mu)$. We consider the case where the
evolution of the brane is slow. This means that in constructing the
effective four--dimensional theory we neglect terms like $(\partial
\phi)^3$.

In addition to the brane positions, we need to include the graviton
zero mode, which can be done by replacing $\eta_{\mu\nu}$ with a
space--time dependent tensor $g_{\mu\nu}(x^\mu)$. Thus, we have
two scalar degrees of freedom, namely the positions of the two
branes which we will denote with $\phi(x^\mu)$ and $\sigma(x^\mu)$,
and the graviton zero mode $g_{\mu\nu}$. 

The effective action is obtained by substituting the metric ansatz in the
5d action, allowing fluctuations of the brane locations. The result to
second order in a derivative expansion reads
\begin{eqnarray}
S_{\rm MSA} &=& \int d^4 x \sqrt{-g_4}\left( f(\phi,\sigma) {\cal R}^{(4)} \right. \nonumber \\
&+& \left. \frac{3}{4}a^2(\phi)\frac{U_B(\phi)}{\kappa_5^2}(\partial \phi)^2 \right.\nonumber \\
&-& \left. \frac{3}{4} a^2(\sigma)\frac{U_B}{\kappa_5^2}(\sigma)(\partial \sigma)^2 \right) .
\end{eqnarray}
where the effective gravitational constant is
\begin{equation}
f(\phi,\sigma) = \frac{1}{\kappa_5^2} \int^{\sigma}_{\phi} dz a^2 (z).
\end{equation}
It is convenient to redefine the moduli fields
\begin{eqnarray}
\tilde \phi^2 &=& \left(1 - 4k\alpha^2 \phi\right)^{2\beta}, \label{posia1}\\
\tilde \sigma^2 &=& \left(1-4k\alpha^2 \sigma\right)^{2\beta} \label{posia2},
\end{eqnarray}
with
$
\beta = \frac{2\alpha^2 + 1}{4\alpha^2};
$
and
\begin{eqnarray}
\tilde \phi &=& Q \cosh R, \label{posib1} \\
\tilde \sigma &=& Q \sinh R \label{posib2}.
\end{eqnarray}
This diagonalises the kinetic terms of the moduli.

The gravitational coupling  can be made constant in the Einstein frame where
\begin{equation}
\tilde g_{\mu\nu} = Q^2 g_{\mu\nu}.
\end{equation}
leading to the effective action
\begin{eqnarray}
S_{\rm EF} &=& \frac{1}{2k\kappa^2_5(2\alpha^2 + 1)}
\int d^4x \sqrt{-g}\left[ {\cal R} \right. \nonumber \\
&-& \left. \frac{12\alpha^2}{1+2\alpha^2}\frac{(\partial Q)^2}{Q^2} \right. \nonumber\\
&-& \left. \frac{6}{2\alpha^2 + 1}(\partial R)^2\right].
\end{eqnarray}
where the gravitational constant is
\begin{equation}
16\pi G = 2k\kappa_5^2 (1+2\alpha^2).
\end{equation}
Notice that the two moduli fields $\phi=\ln Q$ and $R$ are massless fields.
There are two special points in the moduli space. When $R$ vanishes, the
second brane hits the bulk singularity while $Q=0$ corresponds to the
collision of the two branes.
We will see that these special points play a particular in the dynamics
of the brane system.

Let us now discuss the coupling to matter. Matter couples to the induced
metric on the branes. As the bulk is warped the coupling to matter on the first and the second brane differ drastically. In the Einstein frame the matter
action reads
\begin{eqnarray}
S_m^{(1)} &=& S_m^{(1)}(\Psi_1,A^2(Q,R)g_{\mu\nu}) \hspace{0.75 cm}{\rm and}\\
S_m^{(2)} &=& S_m^{(2)}(\Psi_2,B^2(Q,R)g_{\mu\nu}),
\end{eqnarray}
where $\psi_{1,2}$ are the the matter fields on the first and the second brane respectively. The coupling constants $A$ and $B$ are given by
\begin{equation}
A=Q^{-\frac{\alpha^2\lambda}{2}}(\cosh R)^{\frac{\lambda}{4}},\ B=Q^{-\frac{\alpha^2\lambda}{2}}(\sinh R)^{\frac{\lambda}{4}}
\end{equation}
where $\lambda=4/(1+2\alpha^2)$.
Notice that when converging to the singularity, the coupling of the second
brane $B$ vanishes. The validity of the moduli approximation and general
low energy dynamics has been investigated in \cite{gonz} and the radion
dynamics in \cite{sam}.

In what follows, we will concentrate on matter on the first brane only.

\section{Local constraints}
Let us first assume that the two fields $Q$ and $R$ have no potential energy.
The case of a potential for the moduli will
be discussed in Section 6. As matter couples to both gravity and the moduli,
one expects strong deviations
from general relativity. In particular, solar system experiments would detect
the presence of massless moduli unless the Eddington parameter
$\gamma -1\approx -2 \theta$ is close enough to one where
\begin{equation}\label{theta}
\theta= \frac{4}{3}\frac{\alpha^2}{1+2\alpha^2}+ \frac{\tanh^2 R}{6(1+2\alpha^2)}.
\end{equation}
More precisely one must impose $\theta \le 10^{-5}$ (a result obtained in \cite{cassini}) 
implying that
\begin{equation}
\alpha\le 10^{-2},\ R\le 0.01
\end{equation}
today. The smallness
of $\alpha$ indicates a strongly warped bulk geometry such as an
Anti--de Sitter space--time (the Randall-Sundrum setup \cite{rs}).  
In the case $\alpha=0$, we can easily interpret the bound on $R$.  Indeed in that case
\begin{equation}
\tanh R = e^{-k(\sigma -\phi)},
\end{equation}
i.e. this is nothing but the exponential of the radion field measuring
the distance between the branes \cite{radiondef}. Thus we find that
gravity experiments require the branes to be sufficiently far apart.
When $\alpha\ne 0$ but small, one way of obtaining a small value of $R$
is for the hidden brane to become close to the would-be singularity
where $a(\sigma)=0$.

Constraints on the parameters also arise from nucleosynthesis. Nucleosynthesis
constrains the effective number of relativistic degrees of freedom at this
epoch. Apart from this, in brane worlds the energy conservation equation implies
\begin{equation}
\rho a^3\ne const. ,
\end{equation}
resulting in a different expansion rate at the time of nucleosynthesis from that given by
general relativity, giving rise to constraints on the parameters.
One finds $\alpha \le 0.1$ and $R \le 0.4$ at the time of
nucleosynthesis.

\section{Cosmological evolution}
We have seen in the previous section that the parameter
$\alpha$ has to be chosen rather small for
the theory to be consistent with observations. Similarly the field $R$
has to be small too. The field $R$ is dynamical and one would
like to know if the cosmological evolution drives the field $R$ to
small values such that it is consistent with the observations today.
Since the moduli fields are coupled to matter, their evolution is influenced
by the presence of matter. Thus it is necessary to consider the field
equations obtained for the system to study the cosmological evolution
of the brane world moduli.

The field equations for a homogeneous and isotropic universe can be
obtained from the action. The Friedmann
equation reads
\begin{eqnarray}\label{Friedmann}
H^2 &=& \frac{8 \pi G}{3} \left(\rho_1 + \rho_2 + V_{\rm eff}
+ W_{\rm eff} \right) \nonumber \\
&+& \frac{2\alpha^2}{1 + 2\alpha^2} \dot\phi^2
+ \frac{1}{1+2\alpha^2} \dot R^2.
\end{eqnarray}
where we have defined $Q = \exp \phi$.
The field equations for $R$ and $\phi$ read
\begin{eqnarray}
\ddot R &+& 3 H \dot R = - 8 \pi G \frac{1+2\alpha^2}{6}\left[
\frac{\partial V_{\rm eff}}{\partial R} +
\frac{\partial W_{\rm eff}}{\partial R} \right. \nonumber \\
&+& \left. \alpha_R^{(1)} (\rho_1 - 3p_1) +
\alpha_R^{(2)} (\rho_2 - 3p_2) \right] \label{Rcos}
\end{eqnarray}
\begin{eqnarray}
\ddot \phi &+& 3 H \dot \phi =
-8 \pi G \frac{1+2\alpha^2}{12 \alpha^2} \left[
\frac{\partial V_{\rm eff}}{\partial \phi} +
\frac{\partial W_{\rm eff}}{\partial \phi} \right. \nonumber \\
&+& \left. \alpha_\phi^{(1)} (\rho_1 - 3p_1) +
\alpha_\phi^{(2)} (\rho_2 - 3p_2) \right].\label{Qcos}
\end{eqnarray}
The coupling parameters are given by
\begin{eqnarray}
\alpha_\phi^{(1)} &=& -\frac{2\alpha^2}{1+2\alpha^2}, \hspace{0.5cm}
\alpha_\phi^{(2)} = -\frac{2\alpha^2}{1+2\alpha^2}, \label{coupling1} \\
\alpha_R^{(1)} &=& \frac{\tanh R}{1+2\alpha^2}, \hspace{0.5cm}
\alpha_R^{(2)} = \frac{(\tanh R)^{-1}}{1+2\alpha^2}. \label{coupling2}
\end{eqnarray}
We have included matter on both branes as well as potentials
$V_{\rm eff}$ and $W_{\rm eff}$.
We now concentrate on the case where matter is only on our brane.

In the radiation--dominated epoch the trace of the energy--momentum tensor
vanishes, so that
$R$ and $\phi$ quickly become constant. The scale factor scales like
$a(t) \propto t^{1/2}$.

In the matter--dominated era the solution to these equations is given by
\begin{eqnarray}
\rho_1 &=& \rho_e\left(\frac{a}{a_e}\right)^{-3-2\alpha^2/3}, \nonumber \\
a &=& a_e\left(\frac{t}{t_e}\right)^{2/3-4\alpha^2/27}
\end{eqnarray}
together with
\begin{eqnarray}
\phi &=& \phi_e+\frac{1}{3}\ln\frac{a}{a_e} \nonumber \\
R &=& R_0\left(\frac{t}{t_e}\right)^{-1/3}
+ R_1\left(\frac{t}{t_e}\right)^{-2/3},
\end{eqnarray}
as soon as $t\gg t_e$. Note that $R$ indeed decays. This implies
that small values of $R$ compatible with gravitational
experiments are favoured by the cosmological evolution. Note, however,
that the size of $R$ in the early universe is constrained by
nucleosynthesis as well as by the CMB anisotropies.
A large discrepancy between the values of $R$ during
nucleosynthesis and now induces a variation of the particle
masses, or equivalently Newton's constant, which is excluded experimentally.

We can solve the equations numerically to study the cosmological evolution
of the brane world moduli. For matter living on the positive tension brane
and with the potentials $V$ and $W$ identically zero we follow the
evolution of the fields during radiation and matter domination. Both
matter and radiation live on the positive tension brane.  The
calculations are made with $\alpha = 0.01$.

\begin{figure}[htb]
\vspace{9pt}
\includegraphics[width=200pt]{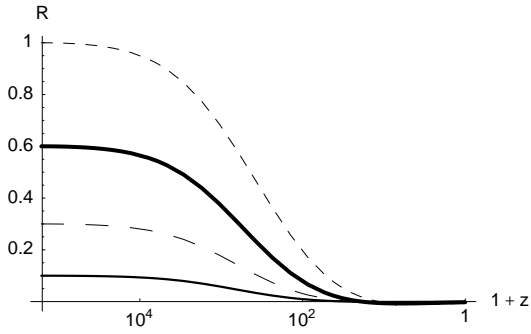}
\caption{The evolution of $R$ with different initial conditions
for the case of radiation and matter on the
positive tension brane and no matter on the negative tension brane.
We find that $R$ is driven towards zero, but if $R$ is too
large at the matter/radiation equality, we find that the
attractor is not efficient enough.}
\end{figure}

\begin{figure}[htb]
\vspace{9pt}
\includegraphics[width=200pt]{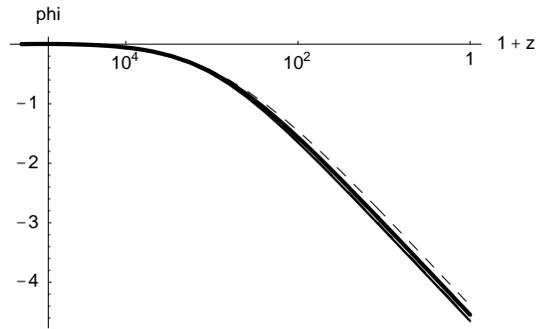}
\caption{The evolution of $\phi$ for different initial
conditions with the same cosmology as in Figure 1.}
\end{figure}

The evolution of $R$ and $\phi$ are shown in fig. 1 and fig. 2,
respectively. One can clearly see, that during radiation domination
both fields are frozen in, because the trace of the energy--momentum
tensor is effectively zero. Soon after matter becomes important,
both fields are forced to evolve due to the non--vanishing trace of
the matter energy--momentum tensor. For the initial conditions we have
chosen the constraint $R<0.1$ can be fullfilled.
One can show that
by putting matter on the negative tension brane as well, the
field $R$ evolves even faster to zero \cite{moduliapprox}. This behaviour
is reminiscent of the attractor solution in scalar--tensor theories \cite{damour}. 
Inflationary dynamics and the consequences for primordial power spectra in this
setup has been studied in \cite{boundary} (see also \cite{boundary2} for a discussion 
of inflation driven by scalar fields on the brane in two-brane systems with bulk 
scalar field).

We now include a potential on the positive tension brane  coming
from supersymmetry breaking effects (see \cite{brax1} and \cite{brax2}). The simplest potential is
obtained by detuning $U_{B}$, so that the potential becomes
\begin{equation}
V =  \frac {6(T-1)k}{\kappa_5^2} e^{\alpha \psi}.
\end{equation}
Here, $T\neq 1$ is a supersymmetry breaking parameter ($\psi$ is the
bulk scalar field). Expressed in terms of $\phi$ and $R$ the effective 
potential on the positive tension brane becomes
\begin{eqnarray}\label{poti}
V_{\rm eff}(\phi,R) &=& \frac{6(T-1)k}{\kappa_5^2} e^{-12\alpha^2\phi/(1+2\alpha^2)} \nonumber \\
&\times& \left(\cosh R \right)^{(4-4\alpha^2)/(1+2\alpha^2)}.
\end{eqnarray}
Notice that for $R$ close to zero, this is nothing but an
exponential model with the field $\phi$ playing the role of a
quintessence field. In the following we set $6k(T-1)$ such that 
the universe starts to accelerate at a redshift around 1. 

\begin{figure}[htb]
\vspace{9pt}
\includegraphics[width=200pt]{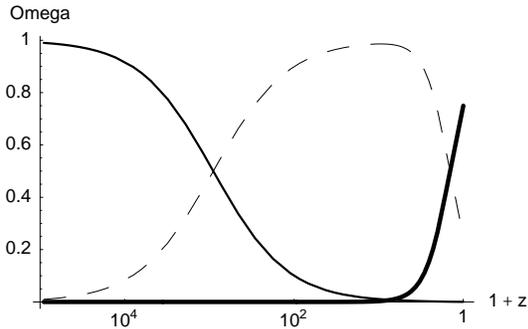}
\caption{Evolution of the density parameter $\Omega_{i}$ as a
function of redshift for radiation, matter and the field $\phi$.
When $\phi$ dominates, the universe is accelerating. Note, that in
order to explain the values for the energy density of dark energy,
one has to fine--tune the parameters of the theory. For these plots we
have set $\alpha=0.01$. The dark matter lives on
the positive tension brane, there is no matter on the negative tension
brane.}
\end{figure}

The evolution of the density parameter is shown in fig. 3.
After the usual matter dominated era, the universe becomes dominated
by the potential energy of the fields and starts to accelerate.

We can study this analytically. The $R$ dependence of the potential for 
small $\alpha$ implies that $R$ is attracted towards zero. Hence, for 
$R$ small the effective potential becomes
\begin{equation}\label{effpot}
V_{\rm eff}(\phi) = \frac{6(T-1)k}{\kappa_5^2} e^{-12\alpha^2 {\phi}/(1+2\alpha^2)}.
\end{equation}
This an exponential potential admitting an attractor with a scale factor
\begin{equation}
a=a_0 t^{\frac{(1+2\alpha^2)}{3\alpha^2}}
\end{equation}
which is accelerating when $\alpha <1$.
The equation of state on the attractor
is given by
\begin{equation}
w=-1+\frac{4}{(1+2\alpha^2)}
\end{equation}

This result has already been obtained using the 5d equations of motion.
Notice that for the Randall-Sundrum case \cite{rs} $\alpha=0$, one gets de Sitter branes.
Another interesting case is $\alpha=-1/\sqrt {12}$ where $w=-5/7$ 
(the supergravity case \cite{brax1}). The five dimensional picture  corresponds to a 
brane moving  at constant speed in the bulk written in terms of conformal 
coordinates.

\section{CMB anisotropies}
In the following we describe how the moduli fields modify
anisotropies in the cosmic microwave background radiation (CMB) \cite{cmbpaper}.
We describe the fields seperately, in order to see their
individual effects. As we have seen, the moduli fields couple to
all matter species on the branes. In the Einstein frame, the
masses of the particles on the branes become moduli-dependent and
therefore the theory is similar to a theory with mass-varying
particles. For the setup considered here, the coupling is {\it
universal}, as long as we assume that the particles on the branes
feel only the induced metric. This means, in particular, that not
only the masses of dark matter particle evolve when the moduli field 
evolve, but also masses of the baryons and neutrinos. In this section
we consider a cosmological constant as dark energy giving rise to
an accelerated universe at the present epoch.

Let us first concentrate on the case $\alpha=0$, so the field $\phi$ decouples from the dynamics. The influence
of $R$ on the CMB anisotropies is plotted in Figure 4.

\begin{figure}[htb]
\vspace{9pt}
\includegraphics[width=150pt,angle=270]{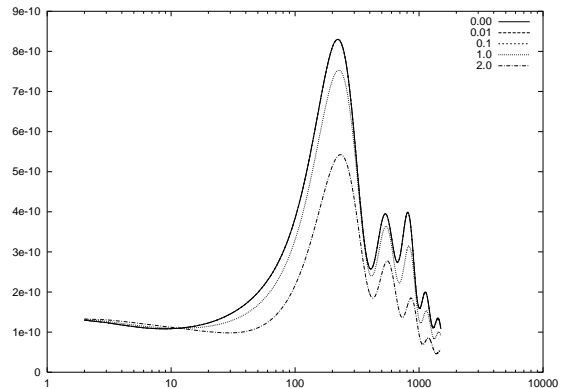}
\caption{COBE normalized CMB anisotropies in the presence of $R$. The cases for different initial values of $R$ (at a redshift
of $10^{10}$) are shown. The smaller the field $R$ is initially, the more the theory behaves like General Relativity.}
\label{fig:largenenough}
\end{figure}

The size of the coupling is set by the initial field value at the
redshift $z\approx 10^{10}$. The larger the field value initially,
the larger is the deviation from General Relativity. The field
starts at a certain value $R_{\rm ini}$ and will, in the matter
dominated epoch,  roll towards zero due to the attractor mechanism
mentioned in Section 4. This will cause the masses of the particle
to vary and will, among other things, modify the evolution of the
metric perturbations. In particular, the gravitational potential is not
time--independent on large scales anymore. This  results in a
larger integrated Sachs-Wolfe (ISW) effect and therefore produce more power on large
angular scales. The normalized spectra have therefore less power
on small angular scales. Since the amount of the different matter
forms at the last scattering surface is changed when compared to
the case $R_{\rm ini} = 0$, the peaks will be shifted and the
relative amplitude of the peaks will modified. These effects can
clearly be seen from Figure 4.

The situation with the field $\phi$ is similar (see Figure 5). The
major difference is that the coupling is constant. This will
modify in particular the ISW at late times: whereas for the field
$R$ the coupling decays in the matter dominated epoch, the
coupling remains constant for the field $\phi$.

\begin{figure}[htb]
\vspace{9pt}
\includegraphics[width=150pt,angle=270]{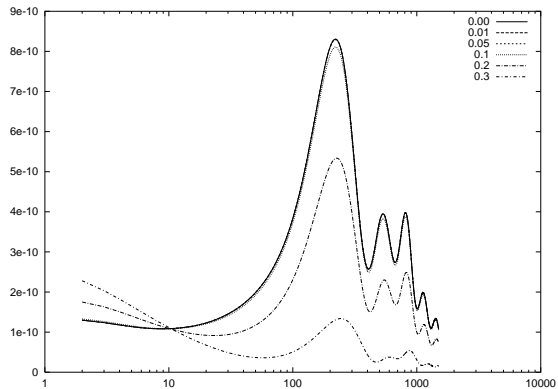}
\caption{COBE normalized CMB anisotropies in the presence of $\phi$. The cases for different initial values of $R$ (at a redshift
of $10^{10}$) are shown.}
\label{fig:largenenough2}
\end{figure}

As it can be seen from Figure 5, already a small coupling $\alpha \approx 0.1$ has a visible impact on the CMB anisotropy spectrum.
To conclude this part, the CMB provides useful constraints on the history of the dynamics of the extra dimension (as encoded in
$R$ and $\phi$). Even imposing the constraints from nucleosynthesis, future observations will put stronger constraints on the
parameter $\alpha$ and $R_{\rm ini}$.

Finally we would like to mention that even if the radion is stabilized after inflation, the stablization mechanism will leave 
an imprint on the primordial perturbations. In \cite{ashcroft} several possibilities were discussed, such as the decay of the radion into radiation 
in the very early universe or a stabilization mechanism without decay. However, since the radion couples to matter on the brane(s), 
during radion stabilization energy will be transfered from the radion to matter and thereby generating entropy perturbations. These 
can leave an imprint in the CMB, which have a different signature than the one discussed above.

\section{The radion as a chameleon}
We have seen that the coupling of the moduli fields leads to modifications
to General Relativity. Since no deviations from General Relativity have
been observed, the fields must have either {\it small} couplings today, or
they must have been stabilized in some way. In this Section we discuss a
particular stabilization mechanism for the radion field.

\subsection{The chameleon mechanism}
The chameleon mechanism provides an alternative mechanism for circumventing
the constraints from local tests of gravity \cite{justin}. According to this idea, the
scalar field(s) acquire(s) a mass which depends on the ambient matter density.
In the cosmos, where the density is minuscule, the mass can be of the
order of the Hubble constant, allowing the field to be rolling on
cosmological time scales. On Earth, however, where the density is many
orders of magnitude higher, the chameleon acquires a mass that is sufficiently
large to satisfy all current experimental bounds on deviations from GR.

\begin{figure}[!ht]
\includegraphics[width=210pt]{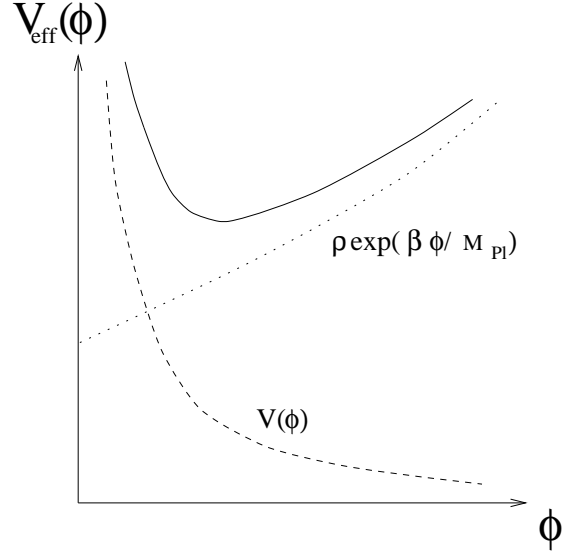}
\caption{The effective potential for the chameleon (solid line) is the sum of the ``bare'' potential,
$V(\phi)$, which is of the runaway form (dashed line), and a density-dependent term (dotted line).
Here we choose the exponential coupling: $A(\phi) = e^{\beta\phi/M_{Pl}}$.}
\label{fig1}
\end{figure}

To see how this works, consider the following general scalar--tensor theory:

\begin{eqnarray}
S &=&\int d^4 x\sqrt{-g} \left\{\frac{M_{Pl}^2R}{2} - \frac{(\partial \phi)^2}{2} \right. \nonumber \\
&-& \left. V(\phi) + {\cal L}_m(\psi_m,A^2(\phi)g_{\mu\nu})\right\}
\end{eqnarray}
where $\phi$ is the chameleon scalar field with scalar potential $V(\phi)$,
assumed to be of the runaway form as in general quintessence models.
See Fig.~\ref{fig1}. Fermion (matter) fields, denoted by $\psi_m$, couple
conformally to the chameleon through the $A^2(\phi)$ dependence of the
matter Lagrangian ${\cal L}_m$.

This conformal coupling leads to an extra term in the Klein-Gordon equation
for the chameleon, as usual proportional to the
trace of the matter stress-tensor:
\begin{equation}
\nabla^2 \phi = V_{,\phi} -  \alpha_\phi T^\mu_\mu\,,
\label{KG1}
\end{equation}
where $\alpha_\phi\equiv \frac{\partial \ln A}{\partial \phi}$.
With the approximation that the matter is well described by a pressureless (non-relativistic) perfect fluid with density $\rho_m$,
this reduces to
\begin{equation}
\nabla^2 \phi = V_{,\phi} + \alpha_\phi\rho_m A(\phi)\,,
\label{KG2}
\end{equation}
where $\rho_m$ is conserved with respect to the Einstein frame metric $g_{\mu\nu}$.

An immediate realization is that the dynamics of $\phi$ are governed by an effective potential, $V_{{\rm eff}}$, which depends
explicitly on $\rho_m$:
\begin{equation}
V_{\rm eff}(\phi)=V(\phi) +\rho_m A(\phi)\,.
\end{equation}
While the ``bare'' potential $V(\phi)$ is of the runaway form, the effective potential
will have a minimum if $A(\phi)$ increases with $\phi$. This is shown in Fig.~\ref{fig1}.
Moreover, the location of this minimum and the mass of small fluctuations, $m^2=V_{,\phi\phi}^{{\rm eff}}$,
both depend on $\rho_m$. In other words, the physical properties of this field vary with the environment,
thus the name chameleon.

It was shown in \cite{justin} that, for constant $\alpha_\phi$, on earth the ``chameleon field'' can have a large mass and hence its interaction is short
($\leq 1mm$, say) scale. This is because the local density is large. In space, however, where the density is lower than on earth, the interaction
can lead to observational consequences. There is a second effect which suppresses the force mediated by the chameleon: For sufficiently large objects,
the $\phi$-force on a test particle is almost entirely due to a thin shell of matter just below the surface of the object, while the matter in the
core of the object contributes negligibly. In other words, only a small fraction of the total mass of the object affects the motion of a test particle
outside. This is the so-called ``thin-shell mechanism''. To illustrate this mechanism in more detail, we derive an approximate solution for the
chameleon for a spherically-symmetric object of radius $R$ and homogeneous density $\rho$. Here we focus on the inverse power-law
potential, $V(\phi) = M^{4+n}/\phi^n$, where $M$ has units of mass, and an exponential coupling of the form $A(\phi)= e^{\beta\phi}$ with
$\beta={\cal O}(1)$. The boundary conditions imposed for this problem are that the solution be non-singular at the origin and that $\phi$
tends to its ambient value, $\phi_0$, far from the object.

For sufficiently large objects, one finds that, within the object, the field assumes a value $\phi_c$ which minimizes the effective
potential, i.e. $V_{,\phi}(\phi_c) + \beta\rho_c e^{\beta\phi_c/M_{Pl}}/M_{Pl} = 0$. This holds everywhere inside the object except
within a thin shell of thickness $\Delta R$ below the surface where the field grows. Outside the object, the profile for
$\phi$ is essentially that of a massive scalar, $\phi \sim \exp(-m_0r)/r$, where $m_0$ is the mass of the chameleon in the ambient medium.

The thickness of the shell is related to $\phi_0$, $\phi_c$,
and the Newtonian potential of the object, $\Phi_N=M/8\pi M_{Pl}^2R$, by
\begin{equation}
\frac{\Delta R}{R} \approx \frac{\phi_\infty-\phi_c}{6\beta
M_{Pl}\Phi_N}\,.
\label{DR}
\end{equation}
The exterior solution can then be written explicitly as~\cite{justin}
\begin{eqnarray}
\phi(r) &\approx& -\left(\frac{\beta}{4\pi M_{Pl}}\right)\left(\frac{3\Delta
R}{R}\right)\frac{M e^{-m_0 (r-R)}}{r} \nonumber \\
&+& \phi_0\,. \label{thinsoln}
\end{eqnarray}

Evidently, this derivation only makes sense if the shell is thin: $\Delta R/R \ll 1$.
Keeping everything else fixed, we see from Eq.~(\ref{DR}) that this is the case for objects with sufficiently large $\Phi_N$.
Then Eq.~(\ref{thinsoln}) says that the correction to Newton's law at short distances is given by $F= (1+\theta) F_N$, where
\begin{equation}
\theta = 2\beta^2 \left(\frac{3\Delta R}{R}\right) \,,
\end{equation}
which is small. Hence a thin shell guarantees a small deviation from Newton's law.

Let us apply this mechanism to gravity experiments.  Fifth force
experiments are usually performed inside a vacuum chamber where
the density is negligibly small. Inside a cavity of radius
$R_{cav}$, explicit calculations~\cite{justin} show that the
chameleon assumes a nearly constant value $\phi_0$, with $\phi_0$
satisfying $m_0 R_{cav} \sim {\cal O}(1)$. That is, the
interaction range of the chameleon-mediated force inside the
cavity is of order of the size of the cavity.

The two test masses used to measure $G$ must have a thin shell,
for otherwise the correction to Newton's constant $G_N$ from the
chameleon-mediated force will be of order unity. Since $G_N$ is
known to an accuracy of $10^{-3}$, they must satisfy
\begin{equation}
\frac{\Delta R}{R} = \frac{\phi_0-\phi_c}{6\beta
M_{Pl}\Phi_N} \;\sim\; 10^{-3}\,.
\end{equation}
where we assume that $\beta=O(1)$.
For typical test masses and cavity of characteristic radii of $\sim 1$~cm and $\sim 1$~m, respectively, this gives
$\phi_0 \sim 10^{-28}\;M_{Pl}$. For the inverse power-law potential, $V(\phi) = M^{4+n}/\phi^n$, with $n\sim {\cal O}(1)$,
this translates into a constraint on $M$:
\begin{equation}
M\sim 10^{-3}\  \hbox{eV}\,.
\label{Mcond}
\end{equation}

This scale is remarkable close to the scale of dark energy, giving
rise to the observed accelerated expansion of the present
universe. The cosmology of this model has been studied in \cite{cosmocham}.

\subsection{Including radion self-interaction}

The question we address now is whether radion self-interaction
will significantly modify the radion evolution and whether the
radion can play the role of dark energy \cite{radion}. To motivate the potential
energy for the radion, we want to avoid the singular point $R=0$
in the evolution (we recall here that the point $R=0$ corresponds
to the situation where  the negative tension brane hits the naked
singularity in the bulk spacetime). We therefore add a potential
energy of the form
\begin{equation}
V(R) = \Lambda^4 R^{-\gamma}.
\end{equation}
leading  to a setup similar to the one discussed above. The major
difference is that the coupling is field dependent. Before
discussing the cosmological consequences of radion
self-interaction, we will investigate whether the thin-shell
mechanism can operate for the radion. In order to simpify our calculations,
we will write the coupling to matter in the form 
\begin{equation}
A(R) \approx 1 + \frac{1}{6}\frac{R^2}{2} +...,
\end{equation}
where we have neglected higher order terms, since the cosmological evolution drives 
the radion naturally to small field values. We found that, imposing the same
boundary conditions as in the chameleon model discussed above, leads to a
force mediated by the radion given by 
\begin{equation}
F_{\phi}=-\frac{\beta^2 R^2_\infty}{4\pi}\frac{m M_{\bullet}}{m_{pl}^2 r^2},
\end{equation}
which is nothing but a correction to Newton's law with (see also Eq. \ref{theta})
\begin{equation}\label{radioncoupling}
\theta=\frac{1}{18} R_\infty^2.
\end{equation}
Therefore, as in the case without a potential for the radion, the strength
of the force mediated by the radion is specified by the cosmological value
of the radion. Since $\theta$ must be small, the (cosmological) field value
must be small.

The cosmological evolution was also studied in \cite{radion} and can be summarized as
follows: During an inflationary epoch, driven by a scalar field confined
on the positive tension brane, the field value is driven towards the minimum.
Assuming a potential energy for the inflaton field $\sigma$ of the form
\begin{equation}
V(\sigma) = \frac{1}{2}m^2 A(R)^2 \sigma^2,
\end{equation}
the field value $R_{\rm min}$ given by
\begin{equation}
R_{\rm min} = \left(\frac{2\gamma\Lambda^4}{3\beta  m^2
 \sigma_{\rm inf}^2}\right)^{1/(\gamma +2)},
\end{equation}
where $\sigma_{\rm inf}$ is the field value of the inflaton field during slow-roll.
Note that the mass of the inflaton field becomes $R$-dependent for the same 
reason that the mass of baryons and dark matter
becomes $R$-dependent.

During the matter and radiation era, the radion converges towards
an attractor. In the radiation era, the attractor is nothing but
the Ratra--Peebles attractor of quintessence. In the matter era,
the attractor does not coincide with the Ratra--Peebles one. In
fact the attractor follows the evolution of the minimum of the
potential lagging behind in such a way that
$R_{\rm Ratra}/R_{min}=const.$ The equation of state of the model is
closer to -1 than in the usual quintessence scenario. This is due
to the friction effect along the quintessence potential induced by
the presence of matter (see figure 7). Now imposing the
gravitational constraints implies that $\Lambda \sim 10^{-3} $ eV
and $\gamma\le 10^{-5}$. This implies that the model must be
extremely close to a cosmological constant throughout the history
of the universe. However, with this fine-tuning the radion can play
the role of dark energy and satisfy local gravitational constraints.
One can avoid the fine--tuning of $\gamma$ by
using the potential $ V(R)= \Lambda ^4 e^{(\frac{M}{R})^n}$. In
this case, the local constraints are satisfied for $n=O(1)$. Again,
$\Lambda \sim 10^{-3} $ eV if the radion is to be a candidate for
dark energy. This tuning is no more than that required by a cosmological
constant.

\section{Conclusions}

Brane-world models have opened up the possibility of testing theories in
extra dimensions through cosmology. At energies below the brane tension
they are well described by a low-energy effective action. We have shown
how this can be computed in the moduli space approximation. For 
brane-world models arising from supergravity there is naturally a bulk
scalar field. The resulting effective action is a bi-scalar-tensor
theory of gravity. This results in corrections to general relativity. 
We have shown how the parameter in the bulk scalar field and the radion 
field today are constrained by solar systems constraints. Similarly brane 
world-moduli change the expansion rate of the universe, resulting in 
constraints on the value of the radion at the time of nucleosynthesis.
The constraints mean that the branes have to be far apart.

The cosmological evolution of brane-world moduli has been studied.
We have shown that the fields evolve in the matter dominated era such
that the branes are driven apart. As a consequence the above constraints
can be satisfied. At late times the bulk scalar field is a natural candidate
for quintessence. We have shown how this field dominates at late times. 

The effective action obtained for the brane-world moduli mean that the
theory is amenable to testing with the CMB. We have computed the CMB 
predictions for both the radion field and the bulk scalar field and 
compared the predictions with those of standard CDM. In both cases there are 
small shifts in the peaks and changes to the ISW effect, when taking into account 
the evolution of the radion and the bulk scalar field. After the above constraints
are imposed these differences with standard CDM are small, but could be
observed in future CMB experiments. 

Finally we have investigated the cosmology of the radion field in more detail.
In the presence of a potential, included to ensure that the negative brane
avoids a bulk singularity, the radion becomes a chameleon field with its
properties depending on the environment, though its couplings to matter
are field dependent. Hence it too can be a dark energy candidate and still satisfy 
solar system constraints, if the constraints on the parameter of the potential 
are imposed. 

We have not covered the evolution of fundamental constants in this paper; a discussion 
can be found in \cite{palma}. Also we have not touched the issue of radion stabilization 
in the early universe. The consequences for cosmological perturbations have been discussed 
in \cite{ashcroft}, see also \cite{kolb}.

\section{Acknowledgements}
We would like to thank our many collaborators, P. Ashcroft, J. Khoury, 
G. Palma, C. Rhodes and A. Weltman, who have made this article possible.
We also thank S. Webster for discussions. This work is supported in part
by PPARC and by a British Council--Alliance exchange grant.

\end{document}